\documentstyle[12pt,aps]{revtex}  
\input{epsfig.sty}

\def\V{{\cal V}}
\def\beq{\begin{equation}}
\def\eqn{\end{equation}\noindent}
\def\r{\vec {\bf r}}

\def\science#1#2#3{{\it Science.}  {\bf #3}, {\it #2}, #1.}
\def\nature#1#2#3{{\it Nature.}  {\bf #3}, {\it #2}, #1.}

\def\pml#1#2#3{{\it Phil. Mag. Lett }  {\bf #3}, {\it #2}, #1.}
\def\pre#1#2#3{{\it Phys. Rev. E }  {\bf #3}, {\it #2}, #1.}

\def\prb#1#2#3{{\it Phys. Rev. B }  {\bf #3}, {\it #2}, #1.}

\def\jcp#1#2#3{{\it J. Chem. Phys. }  {\bf #3}, {\it #2}, #1.}
\def\jcc#1#2#3{{\it J. Comp. Chem. }  {\bf #3}, {\it #2}, #1.}

\def\jpca#1#2#3{{\it J. Phys. Chem. A }  {\bf #3}, {\it #2}, #1.}

\def\cpl#1#2#3{{\it Chem. Phys. Lett. }  {\bf #3}, {\it #2}, #1.}

\def\prl#1#2#3{{\it Phys. Rev. Lett. }  {\bf #3}, {\it #2}, #1.}

\begin{document}
\topmargin=-1.5cm
\textheight=20cm

\title{\bf Global Optimization by Adiabatic Switching}

\author{Jagtar Singh Hunjan and Ramakrishna Ramaswamy}
\address {School of Physical Sciences\\
Jawaharlal Nehru University, New Delhi 110 067, INDIA}
\maketitle

\vspace{1cm}

\begin{abstract}
We apply a recently introduced method for global optimization to
determine the ground state energy and configuration for model metallic
clusters. The global minimum for a given $N$--atom cluster is found by
following the damped dynamics of the $N$ particle system on an
evolving potential energy surface. In this application, the
time dependent interatomic potential interpolates adiabatically between
the Lennard--Jones (LJ) and the Sutton--Chen (SC) forms. Starting with an
ensemble of initial conditions corresponding to the ground state
configuration of the Lennard--Jones cluster, the system asymptotically
reaches the ground state of the Sutton--Chen cluster. We describe
the method and present results for specific cluster size $N=15$,
when the ground state symmetry of LJ$_N$ and SC$_N$ differ.

\vspace{2pt} \hbox to \hsize{\hrulefill}

\end{abstract}
\newpage

\section{Introduction}

Determination of the lowest energy configuration for a 
cluster of $N$ atoms is a nontrivial task 
\cite{walesreview,walesreview2,walestypes,doyec,science2}. 
The complexity arises in part from the exponentially (in $N$) large number 
of minima in the potential energy surface \cite{stillinger}.
Furthermore, the geometry of the potential energy landscape itself can make 
it computationally hard.

The problem is, however, simple to describe: Find the lowest energy minimum
of a $N$ body potential energy surface, $\V(\r)$, 
\beq
\V(\r) = \sum_{i,j} V(r_ij)
\eqn
where $\r$ are the atomic coordinates, and $V(r_{ij})$ is the interatomic
potential of interaction between atoms $i$ and $j$.
For small $N$ one can hope to enumerate all
possible minima and decide the lowest of these, but even for moderate
$N$ and for the simplest $V$ such as the Lennard--Jones (LJ) potential,
this becomes difficult. A case in point is the 38 atom LJ cluster
which has the so--called ``double funnel'' structure; the global minimum,
which has octahedral symmetry, is marginally lower than the first
excited state which has icosahedral symmetry. These were respectively
found by  the basin hopping technique
\cite{wales38} and a genetic algorithm method \cite{deaven,deaven2}.

A number of techniques of global optimization have been applied to 
this problem \cite{walesreview,basinhopping,hartke} and by now
there are extensive compilations of global minima for a number
of different clusters \cite{ccd} , notably those described by
two--body or many--body potentials which are commonly applied in
atomistic simulations. A major difficulty is in 
ensuring that the algorithms reach the global minimum without
being trapped in local minima. One method of overcoming such trapping 
\cite{doyepwales} is by transforming the PES, broadening the 
thermodynamic transitions so as to increase the 
probability of finding the global minimum at temperatures
where the free energy barriers are almost unsurmountable. For example,
addition of a linear term to the PES provides a compressing effect
which has been shown to be successful in locating the true minima in
multiple funneled global structures \cite{doye3}.  Locatelli and Schoen  
\cite{ls} used such transformations to locate the global minimum 
for non-icosahedral clusters.

We have recently proposed a new method of global optimization wherein
{\it time--dependence} is introduced in the potential energy landscape
\cite{hsr}. The evolving landscape is designed in a manner such that 
asymptotically, the potential energy surface develops into the surface 
of interest.  A number of other techniques can be used to follow the 
evolving minima. Applications have been made to determining the 
ground state configurations of simple cluster systems \cite{hsr}.

In the present paper we apply this method to determine the ground state 
configurations and energies of atomic clusters described by the 
many--body Sutton--Chen potential \cite{sc,sc2} by switching from a known 
ground state.  Initially, the interaction is chosen to be the Lennard--Jones 
potential,
\beq
\V_0(\r) = \sum_{ij} V_{LJ}(r_{ij}) = \sum_{ij} 4\epsilon [(\sigma/r_{ij})^{12}
- (\sigma/r_{ij})^{6} ]
\eqn
while the surface of interest is the potential
\beq
\V_f(\r) = \epsilon \sum_i [{1\over 2} \sum_{j\ne i} ({a\over r_{ij}})^n - c\sqrt \rho_i ], ~~~~\rho_i = \sum_{j\ne i} ({a\over r_{ij}})^m
\eqn
One choice for the time dependent potential energy surface is \cite{hsr}
\beq
\V(t) = \V_0(\r) g(t)  + \V_f(\r) h(t)
\label{switch}
\eqn
with $g(t)$ an adiabatically varying switching function that
interpolates between 1 and 0, and $h(t)$ doing the reverse.

In the next section we describe the method as applied to the
problem of ground state energy determination for Sutton--Chen
clusters. Detailed results are presented for one cluster size, while
the more general application and results are indicated in brief.
This is followed by a discussion and summary.

It is a pleasure to dedicate this article to Steve Berry who
has directly and indirectly influenced much of the development in
the area of cluster studies over the past few decades.  We have learned
a lot from him, both in conversation as well as through 
his many articles and reviews \cite{berryreview}.

\section{Adiabatic optimization}

The adiabatic optimization method \cite{hsr}, is a heuristic 
technique for locating minima. The essential idea is as follows.

Time dependence is introduced into the potential energy landscape
directly by the incorporation of slowly varying terms as discussed
in Eq.~(\ref{switch}). A given choice is made for the switching
functions $g(t)$ and $h(t)$, though in practice the choice does
not affect the results greatly. A similar application of 
the adiabatic principle to determine semiclassical ground 
states of multidimensional systems \cite{quant} has noted the
insensitivity of the technique
to the precise form of the switching function,
so long as the induced variation of the potential energy surface
is slow enough.  We note, parenthetically, that
the switching principle has wide applicability, and 
in recent work has been used  in the computation of the 
free--energy of finite clusters \cite{reinhardt}.

Location of the evolving minima can be done by any of a number of 
techniques. The simplest
procedure is to introduce damping into the equations of motion
and allow the system to evolve to a position of rest in a potential
minimum; by starting with an ensemble of initial configurations
and varying the available parameters, a number of minima can be
located, and the putative global minimum can be recognized. 
Elsewhere \cite{hsr} we have suggested the conjugate gradient 
\cite{cg} or simulated annealing (SA) \cite{kirkpatric} 
as other possible methods for locating the minima. It is likely 
that of these, the conjugate gradient technique will be more 
efficient as compared to SA 
though some SA variants \cite{sa} may also provide a
suitable method for following the evolving minima.

The overall procedure can be summarized as follows:
\begin{enumerate}
\item
Take the initial configuration of the $N$ atom cluster to be the
ground state for the LJ$_N$ cluster \cite{ccd}.
\item
Choose some switching function, say $g(t)$. 
Similarly choose $h(t)$,
and the simplest choice, which we make here, is $h(t) = 1-g(t)$.
We have explored a large variety of switching functions and in 
the present application we use $g(t) = \cos^2(3\pi \zeta t) \exp(-\zeta t)$, 
where $\zeta$ is the adiabaticity parameter. 
\item
Perform molecular dynamics simulations for this $N$--particle cluster
with forces deriving from Eq.~(\ref{switch}), with an additional
damping term, namely the equations of motion
\beq
m\ddot{\r_k} + \gamma \dot{\r_k} + {{\partial \V} \over {\partial \r_k}} = 0,~~~~k=1,2\ldots, N,
\eqn
where $\r_k$ is the position vector for the $k$th particle, 
$m$ is its mass and $\gamma$ is the damping constant.
\item
Vary $\zeta$ and $\gamma$, keeping in mind the natural timescales
of the problem. Evolve to a minimum energy configuration, namely when
the particle velocities become zero;  the lowest energy found in an
ensemble of simulations is the ground state energy predicted by
the present method.
\end{enumerate}

\subsection{Results for Sutton--Chen global minima}

Here we attempt to switch from the minimum of the LJ$_N$ system to
the minimum of the SC$_N$ system. Both sets of minima have been
extensively studied earlier and are tabulated in the Cambridge
Cluster Database \cite{ccd}. A point of interest is that for the 
Sutton-Chen
9-6 family of potentials,\cite{sc,sc2} the symmetries of the global minimum
configurations are frequently different from the symmetries of the
Lennard--Jones minima, so that in the adiabatic switching process,
the cluster atoms must also move so as to adopt a different symmetry.

We present detailed results for the  cluster size $N=15$, though we
have applied this technique to larger clusters and obtained results in
agreement with the current standards \cite{ccd}. For the 15 atom LJ
cluster, the ground state has the point group symmetry C$_{2v}$ while
for the SC cluster the symmetry is D$_{6d}$. Shown in Fig.~1 is a plot
of the potential energy versus time for a particular choice of
$g(t), \zeta$ and $\gamma$. Also shown is the effect of 
instantaneous switching,
namely taking the limit $g(t)=0$, where it can be seen that
the system finds the nearest available local minimum from which it
does not move. The time--dependence in the potential effectively
permits the system to explore the multidimensional potential
energy landscape of the SC cluster in an efficient manner. Finding
a local minimum does not trap the system since there is always
kinetic energy {\it until} the adiabatic switch is essentially
over. Inset in the figure is a schematic of the cluster configuration
at different times during the process, showing how the cluster both 
contracts as well as rearranges to eventually reach the minimum of 
the SC surface.

\begin{figure}[!htb]
\centerline{\def\epsfsize#1#2{0.8#1}\epsffile{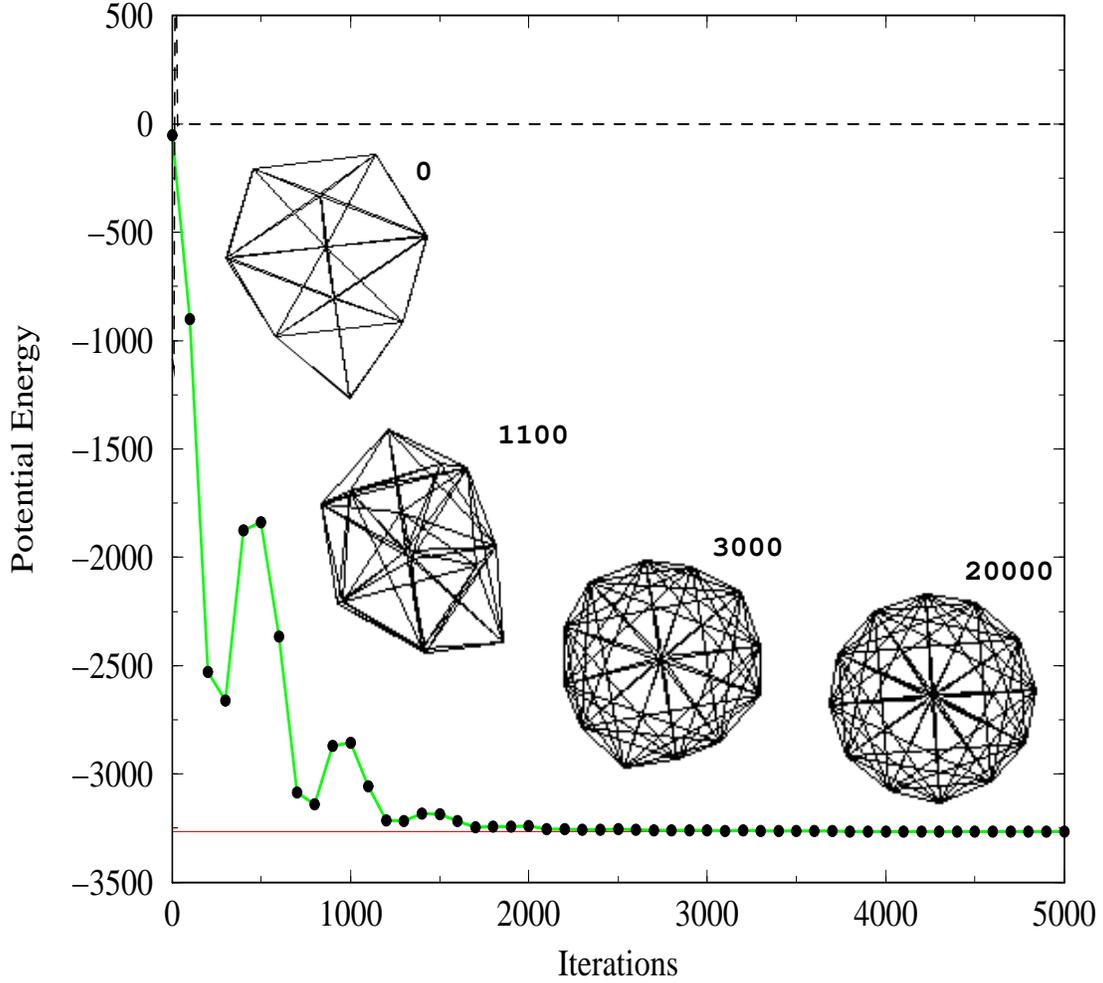}}
\caption
{\label{fig1.eps}
	Plot of the potential energy versus iteration number, switching
from the Lennard Jones potential to the Sutton Chen. The switching function
used is $g(t) = \exp (-\zeta t)\cos^2(3\pi\zeta t)$ with $\zeta = 0.4$
and $\gamma=0.1$. The time step is 0.01 in units natural to the LJ cluster,
for which we also take $\epsilon = \sigma =1$.  At time $t=0$ the cluster
has the $C_{2v}$ symmetry. At different times, as indicated, the cluster
configuration is shown, and asymptotically, the configuration reached is 
the 9--6 Sutton Chen global minimum, with $D_{6d}$ symmetry.
The parameters used for this latter model are taken from [12].
The dashed line shows 
the result of the simulation in the absence of switching,
namely when the LJ potential is {\it suddenly} transformed to the Sutton 
Chen potential.
}
\end{figure}

It should be added that we have performed simulations for a variety
of cluster sizes
and in all cases we find that the procedure successfully finds the
tabulated minima of SC clusters; these are not presented here
since the details are repetitive. As we have emphasized elsewhere
\cite{hsr}, the present method is heuristic, and thus some exploration
of different switching functions, variation in the  adiabaticity and
damping parameters, and indeed the choice of initial potential,
$\V_0(\r)$ is necessary. 

\section{Summary and Discussion}

In this paper we have presented the outline of a general
procedure for global optimization with specific application to
the problem of cluster ground state geometry determination. The
application here, to the determination of the minimum of
model metallic (Sutton-Chen) clusters by adiabatically deforming
the potential energy surface relevant to model rare--gas (Lennard
Jones) clusters is meant to be illustrative rather than exhaustive:
the method introduced here is one of a class of techniques that
employs time--dependence in the potential energy surface to
enhance the exploration of phase space in contrast to other means
of achieving the same objective \cite{basinhopping}.

A multiplicity of techniques is needed to approach hard problems
such as global optimization. Few rigorous results are available,
and application of most techniques is not guaranteed, with few
possible exceptions, to give reliable (or certifiable) results.
The present adiabatic switching method locally solves the optimization
for an evolving surface, and thus mimics other methods of making
large scale excursions in configuration or phase space.

We are presently studying this technique in detail with respect to
the variation of parameters as well as to functional variations.
One of the main issues of concern, and one that we are
addressing in current work, is the relative efficiency
of this method in comparison to other global optimization techniques.
In a number of applications, we find that this method gives
very encouraging results, and permits the determination of fairly
reliable minimum energy configurations for a wide variety of cluster
systems.  The flexibility of choice of a number of starting potentials
including the  free particle case \cite{hsr} as well as the flexible
choice of switching functions and parameters, and finally the flexibility
in the dynamical evolution all combine to suggest that while the method
is heuristic, it holds promise.

\section*{Acknowledgment:} This work is supported by a grant from
the Department of Science and Technology. We thank Subir Sarkar for
discussions.

\end{document}